\documentclass[sigconf,nonacm]{acmart}

\usepackage{float}
\usepackage{color}
\usepackage{url}
\usepackage{pgfplots}
\usepackage{todonotes}
\usepackage{tikz}
\usepackage{mwe}
\usepackage{comment}
\usepackage{listings}
\usepackage{booktabs}
\usepackage{boxedminipage}
\usepackage{svg}
\usepackage{tabu}
\usepackage[justification=centering]{caption}
\usepackage{graphicx}
\usepackage{multirow}
\usepackage{multicol}
\usepackage{arydshln}
\usepackage{threeparttable}
\usepackage{algorithm}
\usepackage{algpseudocode}

\usepackage{subcaption}

\usepackage[inline]{enumitem}

\usepackage{xspace}
\newcommand{\FRONTMATTER}{\textsc{Frontmatter}\xspace}
\newcommand{\GATOR}{\textsc{Gator}\xspace}
\newcommand{\BACKSTAGE}{\textsc{Backstage}\xspace}

\newcommand{\BOOMERANG}{\textsc{Boomerang}\xspace}

\newcommand{\GUI}{GUI\xspace}

\newcommand{\codeid}[1]{\texttt{#1}}
\def\<#1>{\codeid{#1}}

\newenvironment{result}%
{\medskip
\noindent
\let\emph=\textbf\begin{center}
\begin{boxedminipage}{\columnwidth}
    \begin{center}
        \small\em
}%
{
\end{center}
\end{boxedminipage}
\end{center}%
\medskip
}

\usepackage{cleveref}

\begin{document}

\title{
What do all these Buttons do? \\
Statically Mining Android User Interfaces at Scale}

\author{Konstantin Kuznetsov}
\affiliation{
 \institution{CISPA Helmholtz Center for Information Security}
 \city{Saarbr\"ucken}
 \state{Saarland}
 \country{Germany}
}
\email{konstantin.kuznetsov@cispa.saarland}

\author{Chen Fu}
\affiliation{
 \institution{State Key Laboratory of Computer Science, Institute of Software, Chinese Academy of Sciences}
 \institution{University of Chinese Academy of Sciences}
  \city{Beijing}
  \country{China}
}
\email{fchen@ios.ac.cn}

\author{Song Gao}
\affiliation{
 \institution{State Key Laboratory of Computer Science, Institute of Software, Chinese Academy of Sciences}
 \institution{University of Chinese Academy of Sciences}
  \city{Beijing}
  \country{China}
}
\email{gaos@ios.ac.cn}

\author{David N. Jansen}
\affiliation{
 \institution{State Key Laboratory of Computer Science, Institute of Software, Chinese Academy of Sciences}
  \city{Beijing}
  \country{China}
}
\email{dnjansen@ios.ac.cn}

\author{Lijun Zhang}
\affiliation{
 \institution{Institute of Software, Chinese Academy of Sciences}
  \city{Beijing}
  \country{China}
}
\email{zhanglj@ios.ac.cn}

\author{Andreas Zeller}
\affiliation{
 \institution{CISPA Helmholtz Center for Information Security}
 \city{Saarbr\"ucken}
 \state{Saarland}
 \country{Germany}
}
\email{zeller@cispa.saarland}


\begin{abstract}

We introduce \FRONTMATTER: a tool to automatically mine both user interface
models and behavior of Android apps at a large scale with high precision.
Given an app, \FRONTMATTER statically extracts all
declared screens, the user interface elements, their textual and graphical
features, as well as Android APIs invoked by interacting with them.
Executed on tens of thousands of real-world apps, \FRONTMATTER opens the door
for \emph{comprehensive mining of mobile user interfaces}, jumpstarting
empirical research at a large scale, addressing questions such as ``How many
travel apps require registration?'', ``Which apps do not follow accessibility
guidelines?'', ``Does the user interface correspond to the description?'', and
many more. \FRONTMATTER and the mined dataset are
available under an open-source license.

\end{abstract}

\keywords{App mining, user interfaces, Android, app stores, static analysis}

\maketitle

\section{Introduction}
\label{sec:intro}
When designing a user interface, a good practice is to follow the conventions as
set by other user interfaces. But what are these conventions exactly? While
there are myriads of different user interfaces around us, \emph{automated
empirical analysis of these UIs} is a surprisingly difficult task.

The Web, for
instance, sports billions of Web pages. Yet, those Web sites that actually are
accessible for empirical analysis only form the ``surface Web'', with the much
larger ``deep Web'' requiring specific user inputs (such as passwords) to be
accessible. To analyze all pages of a Web application, an automated analysis
thus has to \emph{synthesize} such inputs, which brings all the problems of test
generation. To automatically reach and explore millions of Web applications, one
would have to synthesize interactions for all of these, which is hard---and in
many cases (passwords, internal web pages) simply impossible.

The advent of \emph{app stores} opens an opportunity to \emph{mine}
millions of apps for user interfaces and their properties. Apps can be analyzed
\emph{dynamically} by executing them. However, this approach raises the 
test generation problem, namely to synthesize appropriate
interactions in order to reach and cover the entire user interface. Addressing
this problem can be quite expensive. For instance, the developers of the \emph{Rico} dataset~\cite{Rico17}
spent \$19,200 for crowdsourcing user traces to analyze 9,7K applications.
However, the executable code of apps is also available for
\emph{static analysis.}  This opens the opportunity to analyze millions of apps
for the way they interact with users---and thus \emph{jumpstarting empirical
research on user interfaces at a large scale.} Which apps require a password?
How many of these allow for simple password recovery? Which apps allow to
contact the vendor, and how? Is the concept of a ``shopping cart'' universally
adopted? How do these things change over time?

Many of these questions could be
answered by \emph{statically mining app user interfaces.}
Alas, static analysis of apps and their user interfaces is not exactly easy
either. The challenges are manifold:
\begin{itemize}
\item A user interface analysis should be \emph{complete,} that is, take the
entire user interface into account.
Yet, parts of the user interface may be
constructed \emph{programmatically,} which requires the analysis to identify
(and interpret) the constructing code.
\item A user interface analysis should be \emph{precise,} that is, properly
identify user interface elements, correctly associate them with each other as well as
related code locations.
An imprecise analysis (say, simply extracting all fixed
strings from an executable and its resources) may retrieve all user interface
labels (among many other things), but will not be able to properly group and
organize them.
\item A user interface analysis should \emph{scale,} that is, be applicable to
the large majority of apps as found in app stores. Scalability is an
issue because a precise and complete analysis typically takes time---possibly
too much time for the analysis of hundreds of thousands of apps.
\end{itemize}

In this paper, we present \FRONTMATTER, a complete, precise, and scalable tool
to extract user interface information from Android apps. Given an Android app,
\FRONTMATTER statically identifies all screens reachable within the app; for
every such screen, it extracts the GUI hierarchy of UI elements on the screen,
together with their identifiers, types, user-facing values such as displayed
text and icons, and Android APIs invoked while interacting with them.
This data is extracted from the app manifest and resource files (a description
of all screens and their layout) as well as from the app code itself, in
particular for dynamically defined content. 

To make the analysis \emph{precise}, we adapted the state-of-the-art static
analysis technique by Sp\"ath et al.~\cite{spath2016boomerang} to work on Android apps. Thanks to its flow-,
field-, and context-sensitivity, the \FRONTMATTER static analysis avoids unnecessary
over-approximations: each UI element is placed at the proper screen and matched
with a correct label and a list of triggered API methods. Our implementation also
reduces the number of \emph{points-to} queries needed to resolve particular UI elements.

To accomplish \emph{completeness}, we propose a lightweight procedure to 
augment the code so that the resulting call graph contains more feasible edges, which
are missing otherwise.
At the same time, the respective inter-component
control flow graph (ICFG) is kept concise, which contributes to
\emph{scalability}, as the \emph{alias} and \emph{points-to} analyses may be expensive on large
ICFGs.

\begin{figure*}[t]
\centering
\includegraphics[width=\textwidth]{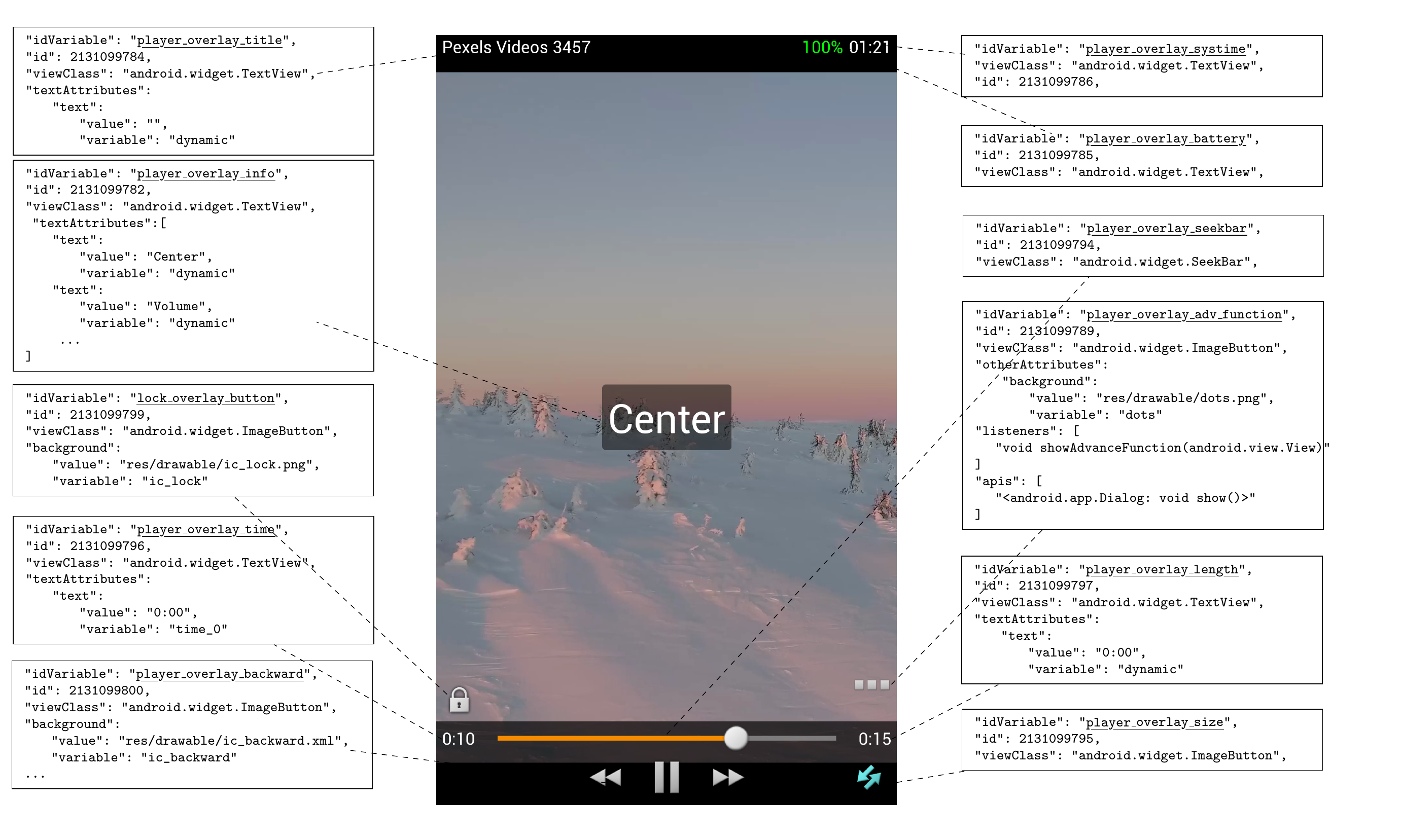}
\caption{UI element data extracted by \FRONTMATTER from a VLC screen (excerpt)}
\label{fig:vlc}
\end{figure*}

As an example of the data \FRONTMATTER provides, consider the well-known VLC
video player app. \Cref{fig:vlc} shows its video-playing screen. For each UI
element, we show an excerpt of the data extracted by \FRONTMATTER. We see that
there are three title elements (title, battery, systime) at the top, a seek
bar with time indicators and control buttons at the bottom. Most prominently
located in the middle is the big \<player\_overlay\_info> text label (showing
\<Center> in the screenshot), which displays various  status messages such as
the current aspect of the video shown. \Cref{fig:vlc} shows only a subset of the
data extracted from the screen. VLC has nine more screens, in total \FRONTMATTER
extracts 214 \GUI elements with 100 attached labels and 16 dialogs from the app.

\FRONTMATTER is not the first tool to analyze Android apps; and it also is not
the first to do so with a focus on user interfaces.  \GATOR~\cite{gator2018} and
\BACKSTAGE~\cite{backstage17} take an app, statically analyze it, and extract
its user interface information, including the association of user interface
elements and related code. However, as we show in this paper, their underlying
analysis is not always precise enough to scale to real-world programs. In
\Cref{fig:vlc}, for instance, \FRONTMATTER identifies 16~possible labels for the
central \<player\_overlay\_info> UI element; these are aspect values such as
\<"Center">, \<"4:3">, \<"16:9">, or status values such as \<"Locked"> or
\<"Unlocked">.  \GATOR also identifies these 16~labels, which is good; however,
due to its context-insensitive nature it also reports six more possible values:
\<"/.nomedia">, \<"+">, \<".">, \<"..">, \<"/..">, and \<"/">. These six
additional values all are file paths or parts thereof (a \<".nomedia"> file in a
folder indicates that this folder should not be indexed by VLC), but never
user-facing on a screen; hence, they are false positives.
While \GATOR over-approximates on this label, \BACKSTAGE under-approximates;
applied to \<player\_overlay\_info>, \BACKSTAGE identifies only one possible
center label \<"4:3">. Likewise, a dynamic execution might not be able to
explore all possible labels, though reporting only feasible ones.

In this paper, we evaluate \FRONTMATTER
against \GATOR and \BACKSTAGE showing that the above observations are not
isolated, but actually common---and while we cannot say that \FRONTMATTER
achieves 100\% precision\footnote{Precisely determining the values a UI element
can take is equivalent to the (undecidable) halting problem; so as with any
program analysis, we must live with imprecision.}, our experiments confirm that
it is sufficiently precise to allow for large scale analysis of user
interfaces.

What can one do with \FRONTMATTER? We have applied \FRONTMATTER on almost
250,000 apps from the AndroZoo database (Google Play apps from 2018, 2019, and
the first half of 2020) and were able to extract user interface data of 70\% of
apps.

The reason why a fraction of the apps from the dataset is considered
has to do with their nature rather than to the faults of our tool:
a lot
of apps are games, cross-platform apps with their own UI builders, and apps
with side code loading. Around half of the initial dataset of 500K apps
consists of games, based on Unity platform, and cross-platform apps (e.g.,
created with Xamarin). As such, they cannot be analyzed with common static analysis
approaches, but only explored dynamically and are therefore filtered out in our study.

This corpus allows us to answer all the questions listed above using
simple database queries. For instance, the query about \textit{Login}
functionality reveals that only 35\% apps in our dataset that have a ``login''
label suggesting to reset the password.  \FRONTMATTER and the extracted data set
are available under open source licenses.

In summary, this paper makes the following contributions:
\begin{enumerate}
  \item \textbf{Context-sensitive reconstruction of user interface
  hierarchies.} \FRONTMATTER is designed from scratch to allow for flow-,
  field-, and context-sensitive extraction of user interface data from Android
  apps. \FRONTMATTER achieves this
  \begin{itemize}
    \item by integrating the \emph{points-to} analysis technique by Sp\"ath et al.~\cite{spath2016boomerang}
    to work with the Android framework; and
    \item through a lightweight procedure to construct the elaborate call graph, preserving the context of callbacks.
  \end{itemize}

  \item \textbf{Associating UI elements and API calls.} The resulting models
  contain not only UI hierarchies, but also \emph{API calls}---reactions of UI
  elements to user interaction, facilitating the analysis of the app's
  behavior.
\end{enumerate}

The remainder of the paper is organized as follows. After discussing the existing
approaches~(\Cref{sec:related}) and the background on Android, its apps, and their user interfaces
(\Cref{sec:background}), we introduce and discuss the analysis techniques of
\FRONTMATTER (\Cref{sec:analysis}). In our evaluation (\Cref{sec:evaluation}),
we compare it to existing approaches~\cite{gator2018,backstage17}. 
\Cref{sec:conclusion} concludes and presents perspectives for future work.

\section{Related Work}
\label{sec:related}%

\subsection{Static Analysis of Android Apps}

\GATOR~\cite{gator2018} is a popular program analysis toolkit for
Android. \GATOR enables and supports static analysis of Android apps, notably
in conjunction with user interaction. \GATOR uses static analysis to model and
extract GUI-related Android objects, their flow through the application, and
their interactions with each other via the Android abstractions. Its focus and
main uses are classical domains of static analysis, including static error
checking, security analysis, test generation, and debugging. This is in contrast
to \FRONTMATTER, which was designed from the ground to extract precise UI models
at scale.

\GATOR is able to extract static models of the user interface from an
app~\cite{rountev2014static}. While its analysis of callback methods to model
window transition graphs is context-sensitive, the reconstruction of window
content is context-insensitive, which leads to some imprecision mentioned in
a couple of subsequent
papers~\cite{yang2015static,wang2016unsoundness,gator2018}. As we will show, \GATOR's precision is not sufficient for extracting models with
accurate label assignment.

The \BACKSTAGE tool~\cite{backstage17} is related to \FRONTMATTER and
\GATOR in that it also aims to analyze apps in conjunction with their user
interface, associating user interface elements with handler and generator
functions. As the name suggests, the aim of \BACKSTAGE is to identify what
happens ``behind the scenes'', i.e. which API calls are triggered when a user
interface element is interacted with. As we show in this work, however,
\BACKSTAGE frequently fails to reconstruct a full UI hierarchy of an app,
missing numerous UI elements and labels. This is in contrast to
\FRONTMATTER, set to reconstruct the complete \GUI model of an app.

\emph{GoalExplorer} by Lai et al.~\cite{lai2019goal} statically models UI screens and screen
transitions in order to guide the dynamic exploration of
an app. Although the authors do not cite \BACKSTAGE, the analysis of the
source code\footnote{https://resess.github.io/PaperAppendices/GoalExplorer/} of
the tool showed that it uses \BACKSTAGE to build an initial UI model. Moreover,
they rely on FlowDroid to build call graphs of all component’s entry points,
whose limitations will be discussed later. GoalExplorer focuses on the
reconstruction of screen transition graphs, while
\FRONTMATTER aims at screen content.

\emph{GUIFetch}~\cite{behrang2018guifetch} searches for apps that are as similar as
possible to the provided sketch of an app. It builds models of app's screens and
screen transitions using a combination of static and dynamic analysis.
In contrast to \FRONTMATTER, though, GUIFetch requires the app source code for its analysis.

\subsection{Dynamic Exploration of Android UIs}

On the Android platform, a number of tools exist to systematically explore
dynamic app behavior. Azim et al. \cite{toolA3E2013} introduced \emph{Targeted
Exploration,} a novel technique that leverages static taint analysis to
facilitate fast yet effective exploration of Android app activities and
Depth-first Exploration, a technique that does not use static analysis, but
instead performs a thorough GUI exploration which results in increased method
coverage.
Bhoraskar et al. \cite{toolbrahmastra2014} presented \emph{Brahmatra,} an app
automation tool addressing the problem that third-party code is usually beyond
the reach of traditional GUI testing tools. It uses static analysis to construct
a page transition graph and discover execution paths to invoke third-party code,
and then performs binary rewriting to ``jump start'' them efficiently.
In contrast to these approaches, \FRONTMATTER uses static analysis, which avoids
the problem of test generation having to cover all functionality.

\subsection{Android UI Datasets}

As for Android UI analysis at scale, Shirazi et al.~\cite{UIAnalysis13}
collected 400 apps from Google Play Store and analyzed the \emph{common design
patterns} of these apps. They estimated the complexity of each app design by
counting the numbers of activities, layout files, and images and computed
descriptive statistics such as the most frequent interface elements and the most
common combinations of widgets.

Alharbi et al.~\cite{UIAnalysis15} applied a data-mining approach to analyze
design pattern changes in Android apps. They tracked 24,436 apps and collected
their versions. After extracting UI elements from these apps, they conducted
differential analyses regarding design pattern changes.

Deka et al.~\cite{Rico17} created the \emph{Rico} dataset
dynamically mining apps and captured view hierarchies, screenshots, and user
interactions with help of crowdsourcing. The \emph{Rico} dataset contains design data from more than 9.7k Android
apps spanning 27 categories. It exposes visual, textual, structural, and
interactive design properties of more than 72k unique UI screens. Based on the
\emph{Rico} dataset, Micallef et al.~\cite{Login18} investigated whether smartphone
apps use login features, and what relationships exist between login features and
apps popularity. 
Liu et al.~\cite{mobiledesignsemantics18} proposed an automatic approach for
annotating mobile UI elements with both structural semantics such as buttons and
functional semantics such as add or search, and computed semantic annotations
for the 72k unique UIs in the \emph{Rico} dataset.
The \emph{ReDraw} data set~\cite{moran2020-tse} is obtained via dynamic analysis, like \emph{Rico}; it was successfully used to prototype GUIs for mobile apps using machine learning techniques.

Being purely dynamic, both the \emph{Rico} and \emph{ReDraw} data sets capture any kind of GUI
including dynamic content and are very precise---\emph{for the executions that have been considered}. However, by construction, their models cannot be complete; it is hard to
cover UI elements that are visible only under certain conditions; and
exploration of an app takes time. Also, neither records API usage.

In contrast to these approaches, \FRONTMATTER is set to be applied automatically
(i.e., without human intervention), statically (i.e., without executing
the app) to the totality of Android app binaries.

\subsection{Mining Web User Interfaces}

\FRONTMATTER is not the first approach to mine UI data. For the Web platform,
\emph{Webzeitgeist}~\cite{designminingweb2013} is a platform for large-scale
design mining comprising a repository of over 100k Web pages and 100 million
design elements, capturing and combining the visual and structural
representations of web pages to compute design features that were used to
power a number of data-driven design tools.  \FRONTMATTER focuses on the
Android domain instead, which allows to take the entire application into
account---in contrast to mining Web applications, where the large majority of
pages is inaccessible for the public.

\subsection{Visual User Interface Mining}

A number of approaches analyze user interfaces \emph{visually}, i.e. from
screen shots and recordings. Pu et al.~\cite{webuivision2017} combined visual
analysis with code mining, increased performance of web page segmentation and
allowed extraction of textual content without the ``costly'' recognition
stage.

Bakaev et al.~\cite{miningwebui2018} applied a vision-based machine learning
approach and built a Web User Interface Visual Analyzer, using image and text
recognition to identify UI elements and labels. In contrast, \FRONTMATTER can
extract the type and labels of visual elements directly from the Android
application.

\subsection{Optimizing Android App analysis}

Besides \GATOR and \BACKSTAGE, recent approaches have been suggested to further
improve static analysis of Android apps. Mirzaei et al. \cite{toolTrimdroid2016}
employed static techniques to identify interaction between GUI widgets and
significantly reduced the number of tests by avoiding the generation of tests
for widgets that do not interact. Zhang et al. \cite{toolCHIME2018} introduced a
launch mode analysis for distinguishing launched instances of an activity class
based on the launch mode specified and employed a launch-mode-aware
context-sensitive activity transition analysis to track activity transitions
context-sensitively, together with an object-sensitive pointer analysis. Both
these optimizations could further improve the precision of \FRONTMATTER, as
well.

\section{Background}
\label{sec:background}

We now provide some background on the Android \GUI, i.e. how
UI elements are declared and how they can be arranged in the layout.

In the Android framework, a single screen containing several graphical user
interface (GUI) elements, such as buttons and text fields, referred to as
widgets, is represented by an \emph{activity} object. Each app typically
contains multiple activities. The layout of the activity is usually declared in
an XML file. This XML file describes the hierarchical structure of the GUI.

All building blocks of Android GUI can be divided into two groups: \emph{views}
and \emph{layouts.} View elements represent specific UI widgets like buttons,
text fields, and images. They are terminal nodes in a GUI tree. Layout elements
serve as containers, which help to organize views in a particular manner, for
instance within a grid or as a scrollable list. Each XML node corresponds to a
certain Android class, either provided by the platform or implemented by a
developer. Thus, the Android framework uses layout files as a \emph{template} to create UI
objects at runtime.

As an example of a layout file, consider \Cref{fig:XMLLayout}. This layout
contains a \textit{TextView} widget which displays text to the user; an
\textit{ImageButton}---a button with an icon; a \textit{ListView} used to show a
list of other widgets; a \textit{fragment} container, which will be substituted
with a layout of SampleFragment during the runtime; and a
\textit{RelativeLayout} container, which is empty now, but can be utilized as an
inflation point for additional UI elements, instantiated in the code.

\begin{lstlisting}[label={fig:XMLLayout},caption={Sample layout.xml Activity layout},escapechar=$,float=hb,numbers=left,
    language=XML, basicstyle=\ttfamily\small, numbersep=-5pt,columns=fullflexible]
  <RelativeLayout>
    <TextView android:id="@+id/info"
      android:text="@string/info"/>     
    <ImageButton android:id="@+id/btn"
      android:src="@drawable/search_src"
      android:onClick="xmlDefinedOnClick"
      android:contentDescription="@string/search"/>
    <ListView android:id="@+id/list"/>
    <fragment android:name="sample.SampleFragment"  $\label{line:xmlfragment}$
      android:id="@+id/fragment"/>
    <RelativeLayout android:id="@+id/container">
  </RelativeLayout>
\end{lstlisting}

Each UI element usually has an
\emph{identifier} (such as \<@+id/info>) that allows the code to
refer to it to activate it or update its content. UI elements also
may have \emph{text} which typically comes as a symbolic label (such
as \<@string/info>) that would be replaced by a string according
to the user's language. Finally, UI elements are tied to
\emph{callbacks}---functions that are invoked when the UI
element is activated. For the \<@+id/btn> button, the \<onClick>
callback is \<xmlDefinedOnClick>: This means that when the button is
clicked, the method \<void xmlDefinedOnClick(View~v)> is invoked,
with \<v> being that button.

\begin{lstlisting}[label={fig:code_oncreate},caption={onCreate method example},
escapechar=$,float=h,numbers=left,language=Java, basicstyle=\small, numbersep=-5pt,columns=fullflexible]
  class SampleActivity extends Activity {
    @Override
    void onCreate(Bundle bundle){
      this.setContentView(R.layout.sample);                             $\label{line:contentres}$
      View info = getActivity().findViewById(R.id.info);
      info.setText("Info");
      List<String> labels = helper.getLabels();
      LinearLayout c = (LinearLayout)this.findViewById(R.id.container); $\label{line:findView}$
      LayoutInflater inflater = LayoutInflater.from(this);
      View container = inflater.inflate(R.layout.new_layout, c, true);  $\label{line:inflate}$
      EditText et = new EditText(this); $\label{line:newView}$
      container.addView(et);
      et.setOnLongClickListener(new View.OnLongClickListener(){         $\label{line:el_callback}$
        @Override 
        public void onLongClick(View v) {
          v.setHint(labels.get(0));
        }
      }
    }
  }
\end{lstlisting}

All activities which are available in an application should be declared in the
app's \emph{manifest file.} Android manages activity instances by invoking
specific \emph{callback methods} that correspond to particular stages of their
life cycle. When the system creates an activity, it invokes the corresponding
\texttt{OnCreate} callback, which should be implemented by the developer.

\Cref{fig:code_oncreate} shows an example code of this method. It is used to
define the layout for the activity's user interface, create and initialize
additional widgets, and bind data to lists. By calling the
\texttt{setContentView(layoutFileId)} method (\Cref{line:contentres}) the system
retrieves a UI hierarchy from the \texttt{layout.xml} file and instantiates necessary Java
classes. Here, \texttt{R.layout.sample} is a unique integer identifier referring
to layout.xml, generated by the compiler. Internally, the \texttt{setContentView}
method fetches a special \texttt{LayoutInflator} object which takes care of
transforming an XML template into a widget instance. This object is also widely
used within the application code to manipulate layouts and their parts. In
\Cref{line:inflate}, the \texttt{LayoutInflator} inflates \textit{new\_layout}
to the GUI hierarchy taking \textit{container} as an attachment point.
Depending
on the signature, the \texttt{LayoutInflator::inflate} method can perform
various actions. Normally, it returns the root \texttt{View} of the created
hierarchy. If the third boolean \emph{attachToRoot} parameter is
provided and \textit{true}, the inflater attaches new widgets to a container.
Otherwise, the view remains dangling and can be manually attached via an
\emph{addView} call. Calls to inflaters may occur in any place of the
application code, where the context is available (either propagated through
method calls or obtained via a system call). Thus, it is important to resolve
\texttt{inflate} invocations in the whole program code.

A widget object provides a number of methods to manipulate its
properties and associate it with text, listeners, or add it to
other widget groups. A reference to an existent \texttt{View} can be
obtained with the \texttt{findViewById} call (\Cref{line:findView}).
Besides using XML files, one can create particular widgets or the
entire layout dynamically in the app code (like
in~\Cref{line:newView}). Usually, listeners of interaction events are
tied to UI elements programmatically. For instance, in \Cref{line:el_callback} the
\texttt{EditText} object is additionally assigned an
\<onLongClick> callback by calling \<View.setOnLongClickListener(l)>
method. These listeners are commonly implemented as anonymous
classes, which may capture the outer variables~(in our example,
labels).
We model callback associations to capture the propagation of
views and control flow.

\begin{lstlisting}[label={fig:code_adapter},caption={ListAdapter handling example},
escapechar=$,float=h,numbers=left,language=Java, basicstyle=\small, numbersep=-5pt,columns=fullflexible]
    class SampleActivity extends Activity {
      void onCreate(Bundle bundle){
        ...
        List<String> items = db.getItems();
        ListView l=(ListView)this.findViewById(R.id.list);
        l.setAdapter(new SearchAdapter(this, items, containerId)); $\label{line:setadapter}$
      }
    }
    class SampleAdapter extends ListAdapter {
      @Override
      public View getView(int position,
                          View convertView, ViewGroup parent) { $\label{line:getview}$
        if (convertView == null) {
            convertView = LayoutInflater.from(context).
                inflate(R.layout.row_items, parent, false);
        }
        return convertView;
      }
    }
\end{lstlisting}

There are Android objects which require special handling of
their layouts. The most frequently used are:

\begin{description}
\item[AdapterView] is a view that displays items loaded into an adapter from
a data-source. For each item to show on a screen, the Android
platform generates a sublayer with the \texttt{getView} method,
(\Cref{fig:code_adapter}),
which should be implemented for each
adapter assigned to the view. Thus, the association of a view element
and an adapter, established by \texttt{setAdapter}
call,
~(\Cref{line:setadapter}),
and adapter layout generation
~(\Cref{line:getview})
should be modeled correctly.

\item[Fragment] is a portion of the user interface that can
combine a single activity to build a multi-pane UI or reuse a module
in multiple activities. There are two different ways to include a
fragment into an activity:
\begin{enumerate*}
\item Declare it in the layout file of the activity
directly (see~\Cref{line:xmlfragment} in~\Cref{fig:XMLLayout} for an
example);
\item Dynamically attach the fragment in the code by means
of the \texttt{FragmentManager} class (~\Cref{line:fragment}in~\Cref{fig:code_fragment}).
\end{enumerate*}

As with adapters, the layout of the fragment is initialized in a
special \textit{onCreateView} callback method---its return statement
should reference a root view container object with an inflated UI tree.
Moreover, it can fetch the execution context and update other
widgets (as in \Cref{line:updateView}).
\end{description}

\begin{lstlisting}[label={fig:code_fragment},caption={Fragment handling example},
escapechar=$,float=h,numbers=left,language=Java, basicstyle=\small, numbersep=-5pt,columns=fullflexible]
    class SampleFragment extends Fragment{
      @Override
      View onCreateView(LayoutInflater li, ViewGroup vg, Bundle b){
        View info = getActivity().findViewById(R.id.info);          $\label{line:updateView}$
        info.setText("Fragment added");
        View view = li.inflate(R.layout.fragment, vg, false);
        return view;
      }
    }
    class SampleActivity extends Activity {
      @Override
      void onCreate(Bundle bundle){
        ...
        FragmentManager fm = this.getSupportFragmentManager();  $\label{line:fragment}$
        FragmentTransaction ft = fm.beginTransaction();
        SampleFragment fragment = new SampleFragment();
        ft.add(R.id.container, fragment);                       $\label{line:add_fragment}$
        ft.commit();
      }
    }
\end{lstlisting}

In summary, in order to construct a precise GUI model of an app, one needs to
correctly identify \emph{creation,} \emph{inflation,} and \emph{attachment of UI
components} (like activities, views, adapters, and fragments) by
determining the control and data flow through respective methods.

\section{Analyzing Android User Interfaces}
\label{sec:analysis}

To properly resolve the several variables involved in assembling the \GUI, one requires a
highly precise \emph{pointer analysis.}  To this end, \FRONTMATTER implements a
static analysis built on top of the SOOT framework~\cite{Soot},
FlowDroid~\cite{Arzt:FlowDroid:PLDI:2014}, and
especially Boomerang~\cite{spath2016boomerang,spath2019context}.

Boomerang is a demand-driven, flow-, field-, and context-sensitive
pointer analysis for Java programs. It computes not only the possible
allocation sites of a given pointer (points-to information), but also
all pointers that can point to those allocation sites (alias
information). 

\subsection{Constructing the Call Graph}

Boomerang's pointer analysis requires an \emph{inter-procedural control
flow graph} based on a corresponding \emph{call graph.} Rather than having a
single entry-point method from which to start the call graph
construction, Android apps combine a large set of entry callbacks,
invoked by the system. Hence, it is necessary to start the analysis
by producing a dummy main method, which models the invocation of these
callbacks.

First, \FRONTMATTER generates an entry point for the analyzed app,
suitable for call graph generation. Then it uses the entry point to
build a control-flow graph, which is used to initialize a
context-sensitive points-to analysis. Finally, \FRONTMATTER identifies
points-of-interest involved in \GUI construction, and performs
resolution of all variables involved.

We utilize a customized \emph{main} method generator from Flowdroid. The
generator aims to accurately model the life-cycle of Android components and the
execution order of related callbacks. However, for our purposes, it still is
incomplete. For instance, it only considers fragments defined in XML layout
templates, thus missing the ones attached dynamically in the code. Hence, the
call graph, generated then by Soot with the SPARK~\cite{lhotak2003scaling}
constructor, lacks edges, rendering parts of the actually executable code
effectively unreachable. To obtain a higher recall along with proper handling of fragments,
we implemented a number of extensions to enrich the final call graph with additional edges.

The call graph assembly procedure encompasses four steps, discussed in the following subsections.

\subsubsection{Step One: Allocation Points of View Objects}

The call graph generated by the SPARK algorithm sometimes misses edges
based on the context of the invoking expression. It is a well known
limitation which happens because there is no object allocation point
for SPARK to use as the starting point for its type
propagation\footnote{E.g. Issues \#439, \#754
\url{https://github.com/Sable/soot/issues/439}}. In order to get these
missing edges, one may, for instance, employ the CHA~\cite{DeanGC95} or RTA~\cite{Bacon:RTA:OOPSLA:96} call graph
construction algorithms, but they are less precise.

This issue of imprecision affects very common code statements. For instance, consider the statement
\begin{center}
\texttt{Button button = (Button)findViewById(id);}
\end{center}
which fetches a UI widget by its id. Although the \texttt{findViewById} method
returns a View reference, it follows the proxy pattern and does not 
explicitly create and return an instance of the View object. As SPARK cannot
identify an object allocation point for this statement, all subsequent method
calls on that object like \texttt{button.setText(text)} are not expanded in the
call graph, thus having no outgoing edges. Hence, all overridden methods (common especially in custom widgets that extend the \texttt{View}
class) become unreachable and thus are excluded from the analysis.

To overcome this problem, we patch the code with \emph{object instantiation
statements.}  At first, we search the code for \texttt{findViewById}
invocations. Next, we reassign a local variable which stores the result of this
call by inserting a new expression statement afterwards. Since this method has
the signature \texttt{View findViewById(int)} in the Jimple
representation\footnote{The intermediate code representation used by
Soot}---i.e.\ it returns an object of type View---it is then usually cast to the
correct type (e.g.\ Button). Where available, instead of insertion we substitute
this cast statement with a new expression. For custom widgets we also insert
their constructor invocation. Later on in our analysis, we distinguish this case
from the standard object allocation.

\subsubsection{Step Two: Asynchronous Execution}

The Android platform provides a couple of classes to perform background
operations asynchronously and publish results on the UI thread. The
most commonly utilized are \texttt{AsyncTask} and \texttt{Handler}.

AsyncTask provides a simple method to handle background threads in
order to update the UI without blocking it by time consuming
operations. Methods of AsyncTask run the code in a separate thread
and then may update the GUI with the results of the execution. However,
they will be not reachable from the dummy main method, since they are
called from the main UI thread by the system. We model this
interaction by transforming asynchronous callback invocations into
\emph{sequential method calls.} To this end, we substitute the
\texttt{execute} call, which activates the task, with four
methods, carrying out the actual work:
\begin{enumerate*}
\item \texttt{onPreExecute()}, which is invoked on the UI thread
right before the execution;
\item the main function
\texttt{doInBackground(Params...)}, invoked afterwards on
the background thread;
\item invoked on the UI thread to display the progress
\texttt{onProgressUpdate(Progress...)}; and finally
\item \texttt{onPostExecute(Result)}, invoked on the UI thread after the
background computation finishes publishing the result.
\end{enumerate*}

In the case of custom workers, a \texttt{Handler} allows communicating
back with the main UI thread from any other background thread. The
message can be sent through the message queue by calling the
\texttt{sendMessage(Message)} method. When the system
receives a Message for a UI thread, it invokes \texttt{handleMessage},
which should be overridden in the app's code. To bind two threads, we
identify sendMessage statements and insert handleMessage calls
afterwards.

\subsubsection{Step Three: Widget Callbacks}

During the construction of the dummy main method, FlowDroid tries to add
invocations of \textit{all} callbacks recognized within an Android
component (e.g., an activity). However, this approach has some limitations.
First, listeners defined outside of activities (such as listeners assigned
in custom methods of the app's classes) are not considered, thus missing
some executed code. More critical, though, is that the identified
callbacks are then invoked inside the dummy main method, so that most
of the parameters required in constructors of the callbacks' objects
are set to null. Hence, FlowDroid does not allow to automatically associate the call
with the specific context, propagate values of particular variables,
and bind them with external fields, which reduces both precision
and recall of the analysis.

To improve callback integration, we patch the code by \emph{injecting the
corresponding callback invoke statements} right after the particular listener is
assigned. For instance, in~\Cref{fig:code_oncreate},~\Cref{line:el_callback} the
\texttt{EditText} widget is attached a long click listener created in-place as
an anonymous class. The Jimple representation of this statement is
\texttt{et.setOnLongClickListener(r1)}, where the variable \texttt{r1}
references previously instantiated \texttt{View.OnLongClickListener} object. We
insert \texttt{r1.onLongClick(et)} statement right after, so that the
\texttt{onLongClick} method receives a) a proper View object \texttt{et} and
b) a variable \texttt{labels}, captured from the outer class.

All in all, we handle 185 Android callbacks (and listeners), of which 75 are from
various UI widgets.

\subsubsection{Step Four: Adapter Callbacks}

In Android, \emph{Adapter} objects provide a special kind of callbacks which
should be handled especially. Adapters function as bridge between the data
and special AdapterView widgets, intended to show this data. They are
responsible for making a View for each item in the data set.
Therefore, we should also model construction of UI elements for at
least one item, inflated into the AdapterView container. This View is
composed inside an \texttt{Adapter::View getView(int position, View
convertView, ViewGroup parent)} method, which should be
implemented by the developer.

To preserve the context, we inject the adapter's \texttt{getView} calls
right after statements with \texttt{setAdapter} method invocation.
(see~\Cref{line:setadapter} in~\Cref{fig:code_adapter}).
After that, all
variables are correctly propagated through the callback.

\subsubsection{Effect of these steps}

All the above steps are performed not on the call graph, but \emph{on the
code right before call graph construction.} Hence, after these steps, we can let Soot build the
call graph as usual, so that it automatically integrates these updates
handling all the peculiarities of method resolution. Finally, we
construct an inter-component control flow graph that we use for
variable resolution in the next step.
According to our preliminary study, the extended call graph contains
\emph{25\% more edges on average} than the one built originally by FlowDroid.

There are also other approaches that address complexities of callback
transition in the Android framework, like
EdgeMiner~\cite{cao2015edgeminer} or Lithium~\cite{perez2017generating}.
While they produce comprehensive callback summaries, it may
overcomplicate the call graph and significantly increase the analysis
time without substantial recall growth. We may consider them in the
future work.

\subsection{Constructing the \GUI Model}

From the call graph, we can now generate the \GUI model.
At first, we analyze methods reachable from
activity callbacks to identify which layouts are associated with that
activity.


For each point-of-interest to analyze, we create a set of backward
search queries for \BOOMERANG to resolve variable values. For example,
for \texttt{this.setContentView(id)} we search backwards in the ICFG for an
allocation site of the \texttt{id} variable to identify which layout
resource is retrieved here. Since we restrict this search with methods
reachable from the activity, we do not need to additionally resolve
the variable \texttt{this}. At the same time, it allows us to correctly
process methods of superclasses.

As we want to analyze an application, but not the Android platform, we
updated \BOOMERANG to skip platform calls. Each time the search
process encounters a method invocation from Android, it jumps over it.

As the next step, we proceed with the resolution of all fragments used
in the app. To this end, we analyze the \emph{onCreateView} method of
each fragment and reconstruct the UI hierarchy generated there. We
apply pointer analysis to the return statement of this method which
contains the reference to the root view of a layout. Afterwards,
\FRONTMATTER identifies which fragments can be added to a particular
activity by tracking \emph{add} and \emph{replace} methods of a
\texttt{FragmentTransaction}.

In the same way, \FRONTMATTER resolves adapters, starting from
\emph{setAdapter} point-of-interest and analyzing \emph{getView} methods.

A \emph{View} object can be either instantiated with the
new-expression statement, or come as a result of an inflation. For
this reason, we extended allocation site definitions of \BOOMERANG to
include not only ordinary basic type constants, final fields, and
new-expression statements, but also specific method invocations
such as \texttt{inflate} of \texttt{LayoutInflator}.
When
\BOOMERANG gives back a points-to sets with a call, \FRONTMATTER makes
another query to determine properties of inflated view.

We also adapted the \BOOMERANG propagation of data-flow facts. In order to
minimize the number of queries per search, we combine queries
and force Boomerang to \emph{transfer facts through invocation statements
regardless of reference types.}
For instance, while performing binding fragments to
activities.
\FRONTMATTER  begins the
resolution with searching for
\texttt{FragmentTransaction::add(container, fragment)} call sites~(see
\Cref{line:add_fragment} of~\Cref{fig:code_fragment}).
At this point,
we aim to identify both the fragment object, which will be added to an
activity, its container, and the target activity. Normally, we would take
this call site as a starting point and query \BOOMERANG for an allocation
statement which was a \texttt{FragmentManager::beginTransaction}
invocation. Next, we would continue propagation with another query which
should have returned a \texttt{getSupportFragmentManager} statement
and finally identified the context of that call. Instead, we tell
\BOOMERANG to pass through the \texttt{beginTransaction} unit and assume that a
data-flow fact is transparently transferred from the
\texttt{FragmentTransaction} to the \texttt{FragmentManager} within one query (although object types are incompatible).

The same strategy applies to the resolution of strings. The updated \BOOMERANG
consumes the \emph{toString} method of a StringBuilder and automatically
determines strings from \texttt{StringBuilder.add(s)} calls.

As \BOOMERANG strives to find \emph{all} allocations and aliases, its
execution time can be quite high. In order to be able to complete the
whole analysis (though, missing some resolutions) we had to set a
timeout of 20~seconds for each query. (We plan to extend \BOOMERANG to
extract partial results.) Finally, \FRONTMATTER aggregates all findings and generates the full \GUI model of the application.

While constructing the \GUI hierarchy, \FRONTMATTER also analyzes
\emph{application behavior}---the reaction of the Android framework triggered in response
to a user interaction with certain UI elements. To this end, \FRONTMATTER
collects all Android APIs which are called by a callback method attached to
each UI element.

\begin{lstlisting}[label={fig:listener},caption={One listener for many buttons},
escapechar=$,float=h,numbers=left,language=Java, basicstyle=\small, numbersep=-5pt,columns=fullflexible]
  public void onClick(View v) {
    Integer vid = v.getId()
    if (R.id.home_button == vid){
        resetState();
        homeButtonClick(v);
    } else if (R.id.next_button == vid){
        nextButtonClick(v);
    }
  }
\end{lstlisting}

It is quite common for developers (for instance, for handling options menus) to
assign a single event listener to multiple UI elements, and make it call different methods
depending on the id of the view that triggered the callback~(\Cref{fig:listener}).
Thus, to properly analyze the behavior of a UI element, it is necessary not only to
correctly associate it with a callback, but also to accurately trace the
execution depending on the view ids involved. Otherwise, the analysis would
over-approximate a lot. Therefore, \FRONTMATTER accounts for the context and
prunes infeasible edges. It walks along the call graph starting from the callback
and takes only proper edges, inspecting whether a condition of \texttt{if} and
\texttt{switch} statements contains a view id constant or a variable initialized
with \texttt{getId} or \texttt{getItemId} methods. The context-sensitivity of
\BOOMERANG and the call graph ensures that the list of collected APIs is
correct.

\subsection{Limitations}
\label{subsec:limitations}

Like any approach for software analysis, \FRONTMATTER has inherent limitations.
As a static analysis tool, it is fundamentally limited by the halting problem;
whether an app will ever show a particular behavior cannot be decided
automatically in all generality, and this of course includes any \GUI features.
\FRONTMATTER thus can overapproximate, reporting \GUI features that are
infeasible in actual executions.

On the other hand, \FRONTMATTER can also \emph{underapproximate,} notably when
code is loaded or generated only when the app executes; such code is not
available at static analysis time. This is the case for apps that embed
\emph{Web} pages; the Web \GUI elements can neither be retrieved nor analyzed or
reported by \FRONTMATTER.

Finally, \FRONTMATTER assumes that the app uses standard Android user interface
elements and concepts. If an app implements its own user interface library,
using only low-level Android input and output primitives, \FRONTMATTER will not
be able to identify or report user interface elements. This is frequently the
case for games as well as for some cross-platform UI libraries; for the latter,
\FRONTMATTER could be adapted to incorporate their concepts.

\section{Evaluation}
\label{sec:evaluation}

We now turn to the evaluation of \FRONTMATTER, focusing on its precision and usefulness compared to the state of the art, as exemplified by \BACKSTAGE and \GATOR. We proceed in four rounds, starting with a single app (VLC) in great detail, and continuing with growing numbers of apps in lesser detail, until we reach the totality of apps in AndroZoo.

\newcommand{\tabincell}[2]{\begin{tabular}{@{}#1@{}}#2\end{tabular}}

\renewcommand{\arraystretch}{1.3}
\setlength\tabcolsep{2pt}
\begin{table*}

\footnotesize
\begin{threeparttable}
\centering
\caption{AndroidBench analysis results produced by: \FRONTMATTER(F), \GATOR(G), and \BACKSTAGE(B)}
\label{tab:eval_gatorbench}
\begin{tabular}{llll|lll|lll|lll|lll|lll}
\toprule
\multirow{2}{*}{APK} &  \multicolumn{3}{c|}{\multirow{2}{*}{Activities:}} &\multicolumn{3}{c|}{\multirow{2}{*}{Identified Views:}} & \multicolumn{3}{c|}{\multirow{2}{*}{View containers}} &\multicolumn{3}{c|}{\multirow{2}{*}{Views with id:}}  & \multicolumn{3}{c|}{\multirow{2}{*}{Widgets with texts}} & \multicolumn{3}{c}{\multirow{2}{*}{Labels per widget}} \\

{} & \multicolumn{3}{c|}{nonempty/empty} & \multicolumn{3}{c|}{correct/incorrect/missing}  & \multicolumn{3}{c|}{/Widgets} & \multicolumn{3}{c|}{total/unique} & \multicolumn{3}{c|}{/Labels} & {} & {} & {} \\
\midrule
                        &(F)    &(G)\tnote{+}    &(B)          &(F)                  &(G)      &(B)        &(F)      &(G)      &(B)       &(F)          &(G)        &(B)       &(F)    &(G)         &(B)    &(F)    &(G)    &(B) \\
BarcodeScanner          &7/2    &8/1    &5/4          &57/6/9         &54/3/12       &50/1/16    &27/36    &24/33    &20/31     & 35/31       & 30/28     & 30/28    &22/30   &26/339     &13/13  &1.36  &13.04  &1.00 \\
Beem                    &10/2   &9/3    &9/3          &86/0/2         &74/0/14       &65/0/23    &36/50    &28/46    &25/40     & 55/46       & 47/40     & 45/38    &32/41   &32/184     &24/25  &1.28  &5.75   &1.04 \\
FBReader                &15/12  &22/5   &11/16        &--             &--            &--         &86/112   &92/112   &37/48     & 155/98      & 165/99    & 49/46    &59/91   &49/336     &15/15  &1.54  &6.86   &1.00 \\
K9                      &22/6   &24/8   &23/6         &--             &--            &--         &188/239  &412/356  &66/94     & 291/126     & 502/120   & 156/101  &131/158 &244/49860  &74/79  &1.21  &204.34 &1.07 \\
KeePassDroid            &12/3   &14/6   &0/15         &--             &--            &--         &28/103   &131/268  &0/0       & 175/88      & 323/105   & 0/0      &85/95   &199/558    &0/0    &1.12  &2.80   &--   \\
Mileage                 &49/1   &49/1   &49/1         &--             &--            &--         &123/161  &313/532  &95/105    & 193/78      & 566/100   & 79/75    &87/103  &235/1477   &41/41  &1.18  &6.29   &1.00 \\
MyTracks                &12/19  &2/29   &12/19        &--             &--            &--         &70/123   &7/7      &39/40     & 137/87      & 8/8       & 40/40    &96/153  &2/2        &25/31  &1.59  &1.00   &1.24 \\
NotePad                 &4/5    &8/5    &4/5          &34/0/4         &21/8/18       &20/0/19    &16/18    &18/18    &9/11      & 30/17       & 29/17     & 11/10    &10/15   &14/115     &4/5    &1.50  &8.21   &1.25 \\
NPR                     &14/0   &14/0   &13/0         &--             &--            &--         &69/95    &3412/5367&62/88     & 113/25      & 4126/58   & 110/25   &31/31   &1680/3764  &24/24  &1.00  &2.24   &1.00 \\
OpenManager             &8/0    &8/0    &8/0          &147/0/0        &132/0/15      &132/0/15   &47/100   &47/85    &47/85     & 74/59       & 62/59     & 62/59    &45/57   &55/360     &38/43  &1.27  &6.55   &1.13 \\
OpenSudoku              &7/3    &7/3    &7/3          &25/0/96        &121/0/0       &25/0/96    &11/14    &31/90    &11/14     & 19/13       & 72/24     & 19/13    &7/8     &61/86      &7/8    &1.14  &1.41   &1.14 \\
SipDroid                &3/8    &3/9    &3/8          &75/0/0         &75/57/0       &31/0/44    &35/40    &43/89    &15/16     & 60/54       & 89/53     & 59/54    &6/11    &43/8104    &4/4    &1.83  &188.47 &1.00 \\
SuperGenpass            &2/1    &2/1    &2/1          &30/0/8         &34/152/4      &30/2/6     &12/18    &70/116   &14/18     & 25/13       & 130/15    & 25/14    &8/14    &10/44      &10/10  &1.75  &4.40   &1.00 \\
TippyTipper             &4/1    &5/1    &4/1          &117/2/2        &117/2/2       &119/0/0    &50/69    &56/83    &50/69     & 36/36       & 48/36     & 36/36    &55/58   &70/80      &54/57  &1.05  &1.14   &1.06 \\
VLC                     &9/1    &9/4    &9/4          &240/4/27       &190/166/77    &101/12/166 &77/167   &109/247  &43/70     & 161/106     & 270/65    & 160/106  &76/122  &86/1625    &37/39  &1.61  &18.90  &1.05 \\
XBMC                    &19/2   &17/5   &17/4         &--             &--            &--         &400/195  &661/493  &144/157   & 321/147     & 925/177   & 258/123  &86/97   &352/13737  &45/53  &1.13  &39.03  &1.18 \\
\bottomrule
\end{tabular}
\begin{tablenotes}
\item[+] In contrast to \FRONTMATTER and \BACKSTAGE, \GATOR also reports activities from dead code, which cannot be accessed by the user.
\item[]\GATOR could not finish the analysis of the Astrid app within 2 hours, while \FRONTMATTER was unable to parse resource files of APV and ConnectBot. Therefore, we excluded these apps from the evaluation.

\end{tablenotes}
\end{threeparttable}
\end{table*}

\subsection{Case Study: VLC Media Player}

In order to determine if \FRONTMATTER can correctly generate a \GUI
model of an app, we examined its performance in detail on a sample application. As a test
subject, we use the VLC media player app introduced in \Cref{sec:intro}, which has been also used to
evaluate \GATOR and \BACKSTAGE.\footnote{We used the version of VLC v.0.0.11 included in the \GATOR replication package. The current version 3.2.6 makes extensive use of \emph{fragments,} which \FRONTMATTER supports, but which cannot be analyzed by \GATOR.}  To obtain a
ground truth, we downloaded the source code of this application and
thoroughly manually investigated the parts responsible for the composition of its UI layouts. Additionally, we ran it in an emulator and recorded each screen state using the UIAutomatorViewer tool from the Android development toolkit.

Comparing \FRONTMATTER and \GATOR against this ground truth, we made two central observations:

\begin{itemize}
\item \textbf{On VLC, both \GATOR and \FRONTMATTER provide a good approximation of the widget set, with \FRONTMATTER being more precise.}  \FRONTMATTER identified the exact set of widgets from the ground truth, while \GATOR inserted 6 copies of widgets in the wrong place of the hierarchy. Besides, it recognized a custom view object, but failed to correctly embed it into the hierarchy, again over-approximating with superfluous UI elements. Since \GATOR does not support fragments, it missed 7 fragments correctly retrieved by \FRONTMATTER.

\item \textbf{\FRONTMATTER is more precise than \GATOR as it comes to precisely assigning textual labels to UI elements,} an observation already discussed in \Cref{sec:intro}. Both
\FRONTMATTER and \GATOR identified most of the labels used in the app.
Even though \GATOR additionally extracted few error messages, it
suffered from the notable over-approximation and assigned the same set
of labels (consisting of 9 elements in one case, and of 18 in the
other) to each encountered \texttt{TextView} widget; among them three strings
are code artifacts and are not declared as UI labels.
In contrast, \FRONTMATTER accurately assigned single labels to the corresponding views, as well as correctly identified 14 labels dynamically assigned to the \texttt{player\_overlay\_info} \texttt{TextView}.
\end{itemize}

These two observations are not limited to VLC alone; as we will see, they apply again and again for Android apps.

\begin{result}
For VLC, \FRONTMATTER constructs a \GUI model of an app\\ with precise association of UI labels.
\end{result}

\subsection{GATOR/AndroidBench Dataset}

In this section, we compare \FRONTMATTER against two contenders, \GATOR and \BACKSTAGE.
For this comparison, we use the
\textit{AndroidBench} benchmark of 19~apps built by the \GATOR authors for the assessment of their tool~\cite{rountev2014static} and used in
follow-up papers; the benchmark is also used to evaluate \BACKSTAGE.
As in the VLC case study, we also determined the ground truth manually for all applications in the data set.

We manually inspected applications in the same way as we assessed VLC: we
downloaded their source code and reconstructed the ground truth. We also used
\texttt{UIautomator} to confirm the layout hierarchy and different properties of
a UI element. To obtain ground truth, for eight out of 16 apps, we fully analyzed
the source code of all activities
To compare hierarchies, we associated each UI node
with its \textit{materialized path} in the tree and then matched these paths.

We ran all three tools on a machine with 64 cores and gave each of them 10GB of
memory. We did not specifically restrict the execution time; still, \GATOR could
not complete the analysis of \emph{Astrid}. Since we used the most recent
version of \GATOR 3.8, its results are different from those reported
in~\cite{rountev2014static}.

Our results are summarized in~\Cref{tab:eval_gatorbench}. In the column `Identified
Views: correct/incorrect/missing' we report the results of our manual analysis:
the number of correctly identified UIs, the number of incorrectly inserted
Views, and the number of missed UI elements (in compared to the ground
truth).

We also report the
following statistics: the number of activities with recognized/undetected
layout; the number of UI elements among which we distinguish view containers
(i.e. view groups) and widgets; the total number of widgets with an id and the
number of unique widgets; the number of widgets with at least one label and the
total number of labels; the average number of captions per widget.

\begin{table*}
\centering
\caption{Analysis results of \FRONTMATTER and \GATOR on extracted models}
\label{tab:androzoo}
\footnotesize
\begin{tabular}{cccccccc}
\hline
& \tabincell{c}{\# Activities\\with/without layout} & \tabincell{c}{\# Elements\\(layouts/widgets)} &\tabincell{c}{ \# Duplicated elements\\/ Unique} &  \tabincell{c}{\# Elements\\per activity} & \tabincell{c}{\# Widgets with texts/\\Total \# of labels} &  \tabincell{c}{\# Labels \\per widgets}\\
\FRONTMATTER & 1511/3313 & 44765 (21851/22914) & 10223/1826& 29.63 & 9332/10191 & 1.10\\
\GATOR       & 1696/2331 & 262376 (75901/186475) & 151093/3573& 154.70 & 45807/1720173 & 37.55\\
\hline
\end{tabular}
\end{table*}

The observations we made for VLC generalize to this set.
\FRONTMATTER correctly models the UI hierarchy, sometimes missing the
content of custom widgets. It precisely associates widget labels, both
of elements with one caption (i.e. buttons) and of text views that store
multiple strings, like status messages.
Occasionally, \GATOR can find more strings, but almost always it overapproximates and assigns
labels to every text widget, with the label-per-widget ratio ranging from 1.18
in TippyTipper to enormous (and highly unrealistic)~202 in K9. (It is hard to
conceive a text label that would display more than 200 different strings encoded
in the program code.) Our analysis of the ground truth supports this observation. 
Besides, \GATOR sometimes inserts a lot of duplicate views into the wrong place.

Moreover, labels returned by \GATOR can contain field and method names,
and even SQL queries. These strings are used internally in the code, rather than displayed to the user.
\BACKSTAGE in general produces more narrow models, but in turn misses elements and text items, sometimes failing to bind an identified layout to an activity.

For four applications, two tools produced almost the same hierarchies with similar label sets.
\FRONTMATTER did not find status messages in the BarcodeScanner app that \GATOR detected. However, \GATOR assigned these strings incorrectly.
For OpenSudoku, \FRONTMATTER could not reconstruct the content of a custom widget. On the other hand, \FRONTMATTER was able to properly process the SipDroid and SuperGenpass apps, for which \GATOR produced very overapproximated model.
A manual investigation of the K9, NPR, and XBMC applications showed that while \FRONTMATTER underapproximated and missed widgets sporadically, \GATOR over-approximated and duplicated a significant portion of elements several times.

\FRONTMATTER offers precision at the cost of running time: it is significantly slower than \GATOR (on average 8x slower on AndroidBench).

\begin{result}
On AndroidBench, \FRONTMATTER retrieves more correct UI~widgets and finds many fewer incorrect labels per widget than \GATOR.
\end{result}

\subsection{AndroZoo Subset}
To test our findings on an even larger set, we ran \FRONTMATTER and \GATOR on a
random subset of 1,000 apps from the AndroZoo database. After removing Unity
based games and apps created with with common cross-platform frameworks, the subset
boiled down to 518 apps.
As in the previous experiment, we ran
both tools on a machine with 64 cores and gave each of them 10GB of
memory, but we restricted the timeout to 10 minutes.

\FRONTMATTER produced models for 403/518 = 78\% of all apps, whereas \GATOR was successful for 385/518 = 74\%. The unsuccessful cases produced timeouts, crashes, and empty models for both tools.

\begin{result}
On the AndroZoo subset, \FRONTMATTER retrieves a much lower number of captions per widget than \GATOR.
\end{result}

Our previous observations also hold for this dataset: \GATOR retrieves more UI elements (see~\Cref{tab:androzoo}).
However, \GATOR produced two outliers (for the apps \emph{com.bibliocommons.mysapl} and \emph{com.directworks.dualapps}) with more than 70,000 elements in the
hierarchy. They account for almost half of the total number of detected
elements. A brief manual examination showed that these apps are unlikely to
have so many widgets. Moreover, we counted UI elements that occur
more than once inside a particular activity of an app (column
\emph{Duplicated elements}). Even though the Android system does not
prohibit widget reuse, certain views are rarely inflated several
times, for \GATOR the ratio of the total number of such elements to
the unique ones is more than~40. Along with the extreme number of labels
per widget~(37.55), this evidence confirms that \GATOR considerably overapproximates.
In contrast, \FRONTMATTER detected fewer widgets, but with higher precision per label.

\subsection{Mining at Scale}
\label{subsec:miningscale}
Finally, we apply \FRONTMATTER on a large scale and construct a
database containing tens of thousands of UI models. To this end, we
crawled AndroZoo for all latest versions of applications downloaded
from the Google Play Store in the first half of 2019. In total we got
72,450 apks. We limit the running time for each app to 10 minutes.
Besides, we set a timeout for each \BOOMERANG query to 20 seconds.

We expect that our analysis is not suitable for some applications:
games usually implement GUI via drawing on canvas, and cross-platform
apps are based on specific frameworks like Mono or Xamarin, which
implement their own way of building an app extensively using
reflection.
For 37,991 applications \FRONTMATTER reported non-empty
results. The analysis of 7180 apps could not finish in time. Other 538
applications crashed Soot with various exceptions. The results of
26741 apks contain a list of activities only. In these cases
\FRONTMATTER could not relate activities to appropriate layouts.

Most of the undetected associations are induced by limitations covered
in~\Cref{subsec:limitations}. For instance, a lot of apps use special
libraries to build the \GUI, such as the Data Binding Library by
Google or Litho by Facebook, which provide custom UI binding
procedures. Besides, the high level of precision can become a
bottleneck. We found that for complex applications \BOOMERANG queries 
during layout binding were too slow to meet the timeout.
With increased running time \FRONTMATTER is able to extract
the model.

\begin{result}
Applied on AndroZoo, \FRONTMATTER extracts rich UI models for 70\% of the apps.
\end{result}

\section{Use Cases}
\label{sec:use-cases}
We have aggregated all data collected in~\Cref{subsec:miningscale} in a
SQLite database, allowing us to easily query the \GUI dataset.

With simple SQL queries one can get insights into various
\GUI features of the dataset. For instance, the query

\begin{lstlisting}[escapechar=$,numbers=left,
    language=SQL, basicstyle=\ttfamily\footnotesize, showstringspaces=false,
    columns=fullflexible, ,numbers=none]
       SELECT DISTINCT pkg, version FROM app-widgets 
                             WHERE (label LIKE '%fingerprint%')
\end{lstlisting}

\noindent lists apps which mention a \emph{fingerprint} login in their interface.
Our database contains 31 such applications.
\pagebreak

Another related query 
\begin{lstlisting}[escapechar=$,numbers=left,
    language=SQL, basicstyle=\ttfamily\footnotesize, showstringspaces=false,
    columns=fullflexible, ,numbers=none]
SELECT COUNT (DISTINCT pkgs from apps_with_login_view)
WHERE (label like '%forget%password%')
     OR(label like '%forgot%password%')
     OR(label like '%reset%password%')
     OR(label like '%recover%password%')
\end{lstlisting}

\noindent
reveals that 45\% of apps which have a login button also have a means to request a lost or forgotten password.

One may be interested how many apps actively ask for (positive) user ratings---706 apps ask to `rate us' according to this query:
\begin{lstlisting}[escapechar=$,numbers=left,
    language=SQL, basicstyle=\ttfamily\footnotesize, showstringspaces=false,
    columns=fullflexible, ,numbers=none]
SELECT  COUNT(DISTINCT pkg)
FROM appwidgets WHERE (label like '%rate us%')
    OR (label like '%rate 5%')
    OR (label like '%rate our%') OR (label like '%rate the app%')
    OR (label like '%rate this app%') OR (label like '%rate it%')

\end{lstlisting}

The dataset also contains information about the drawables used in apps, such as the color or an icon name:
\begin{lstlisting}[escapechar=$,numbers=left,
    language=SQL, basicstyle=\ttfamily\footnotesize, showstringspaces=false,
    columns=fullflexible, ,numbers=none]
        SELECT drawable, COUNT(drawable) AS v_o FROM pkgs
              WHERE d.drawable_type like '%background%'
              GROUP BY drawable ORDER BY v_o DESC
\end{lstlisting}

\noindent
shows that the most frequently used background colors are white, black, and gray, which is not surprising.

All these basic features can be combined for further refinement, and of course correlated with additional features obtained from app store analysis (``Rating requests are most prominent in app category $X$'') or program analysis (``Applications that have a label $X$ also use API $Y$'').



\section{Conclusion and Future Work}
\label{sec:conclusion}

We have introduced \FRONTMATTER, a tool to automatically extract user interfaces
of Android applications at a large scale with high precision. Given an app,
\FRONTMATTER statically extracts all accessible screens, the user interface
elements, as well as their textual and graphical features, increasing precision
over the state of the art. Executed on tens of thousands of real-world apps,
\FRONTMATTER opens the door for \emph{comprehensive mining of mobile user
interfaces}, jumpstarting empirical research at a large scale.
Our own work will focus on deeper
analysis of the mined \FRONTMATTER data:

\begin{itemize}
\item \textbf{Do interface texts correspond to app descriptions?}  The set of
user interface labels form linguistic \emph{concepts,} as do the set of words in
app descriptions~\cite{chabada}; comparing these might reveal mismatches between user interfaces and functionality.

\item \textbf{Which UI features are common for specific categories? Which
ones are uncommon?}  ``Travel'' apps would typically sport user interfaces to
query locations, book rooms, or book trips; for ``Music'' apps, we would expect
interfaces to play, well, music. But would there be apps that sport unique
features found nowhere else?

\item \textbf{Are there specific \emph{user interface patterns} for specific
features?}  Which are the features typically provided by product pages, by
registration pages, by query pages? Many applications require passwords to
login. Most of these would also offer to recover passwords, or to register as a
new user. Are there apps which require login and which do not provide such
features? Should they?

\item \textbf{User interfaces and popularity.}  Would especially popular apps
sport user interface elements that others do not? What is it in their user
interfaces that makes popular apps special? Would ``polished'' interfaces
manifest themselves, and how?

\item \textbf{Halls of fame, halls of shame.}  Just like patterns for good user
interfaces, would there be \emph{anti-patterns}, too? What are the most common
mistakes to avoid, and can one detect them from the extracted data? What is the
worst interface ever?

\end{itemize}

To foster replication and extension of our research, \FRONTMATTER and the mined user interface data are available under open source and open data licenses at

\begin{center}
\url{https://bit.ly/3knQHc9}
\end{center}

\bibliographystyle{ACM-Reference-Format}
\bibliography{ref}


\begin{thebibliography}{30}


\ifx \showCODEN    \undefined \def \showCODEN     #1{\unskip}     \fi
\ifx \showDOI      \undefined \def \showDOI       #1{#1}\fi
\ifx \showISBNx    \undefined \def \showISBNx     #1{\unskip}     \fi
\ifx \showISBNxiii \undefined \def \showISBNxiii  #1{\unskip}     \fi
\ifx \showISSN     \undefined \def \showISSN      #1{\unskip}     \fi
\ifx \showLCCN     \undefined \def \showLCCN      #1{\unskip}     \fi
\ifx \shownote     \undefined \def \shownote      #1{#1}          \fi
\ifx \showarticletitle \undefined \def \showarticletitle #1{#1}   \fi
\ifx \showURL      \undefined \def \showURL       {\relax}        \fi
\providecommand\bibfield[2]{#2}
\providecommand\bibinfo[2]{#2}
\providecommand\natexlab[1]{#1}
\providecommand\showeprint[2][]{arXiv:#2}

\bibitem[\protect\citeauthoryear{Alharbi and Yeh}{Alharbi and Yeh}{2015}]%
        {UIAnalysis15}
\bibfield{author}{\bibinfo{person}{Khalid Alharbi} {and} \bibinfo{person}{Tom
  Yeh}.} \bibinfo{year}{2015}\natexlab{}.
\newblock \showarticletitle{Collect, Decompile, Extract, Stats, and Diff:
  Mining Design Pattern Changes in {Android} Apps}. In
  \bibinfo{booktitle}{\emph{Proceedings of the 17th International Conference on
  Human-Computer Interaction with Mobile Devices and Services}} (Copenhagen,
  Denmark) \emph{(\bibinfo{series}{MobileHCI '15})}.
  \bibinfo{publisher}{Association for Computing Machinery},
  \bibinfo{address}{New York, NY, USA}, \bibinfo{pages}{515--524}.
\newblock
\showISBNx{9781450336529}
\urldef\tempurl%
\url{https://doi.org/10.1145/2785830.2785892}
\showDOI{\tempurl}


\bibitem[\protect\citeauthoryear{Arzt, Rasthofer, Fritz, Bodden, Bartel, Klein,
  Le~Traon, Octeau, and McDaniel}{Arzt et~al\mbox{.}}{2014}]%
        {Arzt:FlowDroid:PLDI:2014}
\bibfield{author}{\bibinfo{person}{Steven Arzt}, \bibinfo{person}{Siegfried
  Rasthofer}, \bibinfo{person}{Christian Fritz}, \bibinfo{person}{Eric Bodden},
  \bibinfo{person}{Alexandre Bartel}, \bibinfo{person}{Jacques Klein},
  \bibinfo{person}{Yves Le~Traon}, \bibinfo{person}{Damien Octeau}, {and}
  \bibinfo{person}{Patrick McDaniel}.} \bibinfo{year}{2014}\natexlab{}.
\newblock \showarticletitle{{FlowDroid}: Precise Context, Flow, Field,
  Object-sensitive and Lifecycle-aware Taint Analysis for {Android} Apps}.
  \bibinfo{publisher}{Association for Computing Machinery},
  \bibinfo{address}{New York, NY, USA}, \bibinfo{pages}{259--269}.
\newblock
\showISBNx{978-1-4503-2784-8}
\urldef\tempurl%
\url{https://doi.org/10.1145/2594291.2594299}
\showDOI{\tempurl}


\bibitem[\protect\citeauthoryear{{Avdiienko}, {Kuznetsov}, {Rommelfanger},
  {Rau}, {Gorla}, and {Zeller}}{{Avdiienko} et~al\mbox{.}}{2017}]%
        {backstage17}
\bibfield{author}{\bibinfo{person}{V. {Avdiienko}}, \bibinfo{person}{K.
  {Kuznetsov}}, \bibinfo{person}{I. {Rommelfanger}}, \bibinfo{person}{A.
  {Rau}}, \bibinfo{person}{A. {Gorla}}, {and} \bibinfo{person}{A. {Zeller}}.}
  \bibinfo{year}{2017}\natexlab{}.
\newblock \showarticletitle{Detecting Behavior Anomalies in Graphical User
  Interfaces}. In \bibinfo{booktitle}{\emph{2017 IEEE/ACM 39th International
  Conference on Software Engineering Companion (ICSE-C)}}.
  \bibinfo{pages}{201--203}.
\newblock
\showISSN{null}
\urldef\tempurl%
\url{https://doi.org/10.1109/ICSE-C.2017.130}
\showDOI{\tempurl}


\bibitem[\protect\citeauthoryear{Azim and Neamtiu}{Azim and Neamtiu}{2013}]%
        {toolA3E2013}
\bibfield{author}{\bibinfo{person}{Tanzirul Azim} {and} \bibinfo{person}{Iulian
  Neamtiu}.} \bibinfo{year}{2013}\natexlab{}.
\newblock \showarticletitle{Targeted and depth-first exploration for systematic
  testing of {Android} apps}. In \bibinfo{booktitle}{\emph{Proceedings of the
  2013 ACM SIGPLAN international conference on Object oriented programming
  systems languages \& applications}}. \bibinfo{pages}{641--660}.
\newblock


\bibitem[\protect\citeauthoryear{Bacon and Sweeney}{Bacon and Sweeney}{1996}]%
        {Bacon:RTA:OOPSLA:96}
\bibfield{author}{\bibinfo{person}{David~F. Bacon} {and}
  \bibinfo{person}{Peter~F. Sweeney}.} \bibinfo{year}{1996}\natexlab{}.
\newblock \showarticletitle{Fast Static Analysis of {C++} Virtual Function
  Calls}, \bibfield{editor}{\bibinfo{person}{Lougie Anderson} {and}
  \bibinfo{person}{James Coplien}} (Eds.). \bibinfo{publisher}{{ACM}},
  \bibinfo{pages}{324--341}.
\newblock
\showISBNx{0-89791-788-X}
\urldef\tempurl%
\url{https://doi.org/10.1145/236337.236371}
\showDOI{\tempurl}


\bibitem[\protect\citeauthoryear{Bakaev, Heil, Khvorostov, and Gaedke}{Bakaev
  et~al\mbox{.}}{2018}]%
        {miningwebui2018}
\bibfield{author}{\bibinfo{person}{Maxim Bakaev}, \bibinfo{person}{Sebastian
  Heil}, \bibinfo{person}{Vladimir Khvorostov}, {and} \bibinfo{person}{Martin
  Gaedke}.} \bibinfo{year}{2018}\natexlab{}.
\newblock \showarticletitle{{HCI} Vision for automated analysis and mining of
  web user interfaces}. In \bibinfo{booktitle}{\emph{International Conference
  on Web Engineering}}. Springer, \bibinfo{pages}{136--144}.
\newblock


\bibitem[\protect\citeauthoryear{Behrang, Reiss, and Orso}{Behrang
  et~al\mbox{.}}{2018}]%
        {behrang2018guifetch}
\bibfield{author}{\bibinfo{person}{Farnaz Behrang}, \bibinfo{person}{Steven~P
  Reiss}, {and} \bibinfo{person}{Alessandro Orso}.}
  \bibinfo{year}{2018}\natexlab{}.
\newblock \showarticletitle{{GUIfetch}: supporting app design and development
  through {GUI} search}. In \bibinfo{booktitle}{\emph{Proceedings of the 5th
  International Conference on Mobile Software Engineering and Systems}}.
  \bibinfo{pages}{236--246}.
\newblock


\bibitem[\protect\citeauthoryear{Bhoraskar, Han, Jeon, Azim, Chen, Jung, Nath,
  Wang, and Wetherall}{Bhoraskar et~al\mbox{.}}{2014}]%
        {toolbrahmastra2014}
\bibfield{author}{\bibinfo{person}{Ravi Bhoraskar}, \bibinfo{person}{Seungyeop
  Han}, \bibinfo{person}{Jinseong Jeon}, \bibinfo{person}{Tanzirul Azim},
  \bibinfo{person}{Shuo Chen}, \bibinfo{person}{Jaeyeon Jung},
  \bibinfo{person}{Suman Nath}, \bibinfo{person}{Rui Wang}, {and}
  \bibinfo{person}{David Wetherall}.} \bibinfo{year}{2014}\natexlab{}.
\newblock \showarticletitle{Brahmastra: Driving apps to test the security of
  third-party components}. In \bibinfo{booktitle}{\emph{23rd $\{$USENIX$\}$
  Security Symposium ($\{$USENIX$\}$ Security 14)}}.
  \bibinfo{pages}{1021--1036}.
\newblock


\bibitem[\protect\citeauthoryear{Cao, Fratantonio, Bianchi, Egele, Kruegel,
  Vigna, and Chen}{Cao et~al\mbox{.}}{2015}]%
        {cao2015edgeminer}
\bibfield{author}{\bibinfo{person}{Yinzhi Cao}, \bibinfo{person}{Yanick
  Fratantonio}, \bibinfo{person}{Antonio Bianchi}, \bibinfo{person}{Manuel
  Egele}, \bibinfo{person}{Christopher Kruegel}, \bibinfo{person}{Giovanni
  Vigna}, {and} \bibinfo{person}{Yan Chen}.} \bibinfo{year}{2015}\natexlab{}.
\newblock \showarticletitle{{EdgeMiner}: Automatically Detecting Implicit
  Control Flow Transitions through the {Android} Framework.}. In
  \bibinfo{booktitle}{\emph{NDSS}}.
\newblock


\bibitem[\protect\citeauthoryear{Dean, Grove, and Chambers}{Dean
  et~al\mbox{.}}{1995}]%
        {DeanGC95}
\bibfield{author}{\bibinfo{person}{Jeffrey Dean}, \bibinfo{person}{David
  Grove}, {and} \bibinfo{person}{Craig Chambers}.}
  \bibinfo{year}{1995}\natexlab{}.
\newblock \showarticletitle{Optimization of Object-Oriented Programs Using
  Static Class Hierarchy Analysis} \emph{(\bibinfo{series}{Lecture Notes in
  Computer Science})}, \bibfield{editor}{\bibinfo{person}{Walter~G. Olthoff}}
  (Ed.), Vol.~\bibinfo{volume}{952}. \bibinfo{publisher}{Springer},
  \bibinfo{pages}{77--101}.
\newblock
\showISBNx{3-540-60160-0}
\urldef\tempurl%
\url{https://doi.org/10.1007/3-540-49538-X}
\showDOI{\tempurl}


\bibitem[\protect\citeauthoryear{Deka, Huang, Franzen, Hibschman, Afergan, Li,
  Nichols, and Kumar}{Deka et~al\mbox{.}}{2017}]%
        {Rico17}
\bibfield{author}{\bibinfo{person}{Biplab Deka}, \bibinfo{person}{Zifeng
  Huang}, \bibinfo{person}{Chad Franzen}, \bibinfo{person}{Joshua Hibschman},
  \bibinfo{person}{Daniel Afergan}, \bibinfo{person}{Yang Li},
  \bibinfo{person}{Jeffrey Nichols}, {and} \bibinfo{person}{Ranjitha Kumar}.}
  \bibinfo{year}{2017}\natexlab{}.
\newblock \showarticletitle{Rico: A Mobile App Dataset for Building Data-Driven
  Design Applications}. In \bibinfo{booktitle}{\emph{Proceedings of the 30th
  Annual ACM Symposium on User Interface Software and Technology}} (Qu\'{e}bec
  City, QC, Canada) \emph{(\bibinfo{series}{UIST '17})}.
  \bibinfo{publisher}{Association for Computing Machinery},
  \bibinfo{address}{New York, NY, USA}, \bibinfo{pages}{845--854}.
\newblock
\showISBNx{9781450349819}
\urldef\tempurl%
\url{https://doi.org/10.1145/3126594.3126651}
\showDOI{\tempurl}


\bibitem[\protect\citeauthoryear{Gorla, Tavecchia, Gross, and Zeller}{Gorla
  et~al\mbox{.}}{2014}]%
        {chabada}
\bibfield{author}{\bibinfo{person}{Alessandra Gorla}, \bibinfo{person}{Ilaria
  Tavecchia}, \bibinfo{person}{Florian Gross}, {and} \bibinfo{person}{Andreas
  Zeller}.} \bibinfo{year}{2014}\natexlab{}.
\newblock \showarticletitle{Checking App Behavior against App Descriptions}. In
  \bibinfo{booktitle}{\emph{Proceedings of the 36th International Conference on
  Software Engineering}} (Hyderabad, India) \emph{(\bibinfo{series}{ICSE
  2014})}. \bibinfo{publisher}{Association for Computing Machinery},
  \bibinfo{address}{New York, NY, USA}, \bibinfo{pages}{1025--1035}.
\newblock
\showISBNx{9781450327565}
\urldef\tempurl%
\url{https://doi.org/10.1145/2568225.2568276}
\showDOI{\tempurl}


\bibitem[\protect\citeauthoryear{Kumar, Satyanarayan, Torres, Lim, Ahmad,
  Klemmer, and Talton}{Kumar et~al\mbox{.}}{2013}]%
        {designminingweb2013}
\bibfield{author}{\bibinfo{person}{Ranjitha Kumar}, \bibinfo{person}{Arvind
  Satyanarayan}, \bibinfo{person}{Cesar Torres}, \bibinfo{person}{Maxine Lim},
  \bibinfo{person}{Salman Ahmad}, \bibinfo{person}{Scott~R Klemmer}, {and}
  \bibinfo{person}{Jerry~O Talton}.} \bibinfo{year}{2013}\natexlab{}.
\newblock \showarticletitle{Webzeitgeist: design mining the web}. In
  \bibinfo{booktitle}{\emph{Proceedings of the SIGCHI Conference on Human
  Factors in Computing Systems}}. ACM, \bibinfo{pages}{3083--3092}.
\newblock


\bibitem[\protect\citeauthoryear{Lai and Rubin}{Lai and Rubin}{2019}]%
        {lai2019goal}
\bibfield{author}{\bibinfo{person}{Duling Lai} {and} \bibinfo{person}{Julia
  Rubin}.} \bibinfo{year}{2019}\natexlab{}.
\newblock \showarticletitle{Goal-driven exploration for Android applications}.
  In \bibinfo{booktitle}{\emph{2019 34th IEEE/ACM International Conference on
  Automated Software Engineering (ASE)}}. IEEE, \bibinfo{pages}{115--127}.
\newblock


\bibitem[\protect\citeauthoryear{Lhot{\'a}k and Hendren}{Lhot{\'a}k and
  Hendren}{2003}]%
        {lhotak2003scaling}
\bibfield{author}{\bibinfo{person}{Ond{\v{r}}ej Lhot{\'a}k} {and}
  \bibinfo{person}{Laurie Hendren}.} \bibinfo{year}{2003}\natexlab{}.
\newblock \showarticletitle{Scaling {Java} points-to analysis using {SPARK}}.
  In \bibinfo{booktitle}{\emph{International Conference on Compiler
  Construction}}. Springer, \bibinfo{pages}{153--169}.
\newblock


\bibitem[\protect\citeauthoryear{Liu, Craft, Situ, Yumer, Mech, and Kumar}{Liu
  et~al\mbox{.}}{2018}]%
        {mobiledesignsemantics18}
\bibfield{author}{\bibinfo{person}{Thomas~F. Liu}, \bibinfo{person}{Mark
  Craft}, \bibinfo{person}{Jason Situ}, \bibinfo{person}{Ersin Yumer},
  \bibinfo{person}{Radomir Mech}, {and} \bibinfo{person}{Ranjitha Kumar}.}
  \bibinfo{year}{2018}\natexlab{}.
\newblock \showarticletitle{Learning Design Semantics for Mobile Apps}. In
  \bibinfo{booktitle}{\emph{The 31st Annual ACM Symposium on User Interface
  Software and Technology}} (Berlin, Germany) \emph{(\bibinfo{series}{UIST
  '18})}. \bibinfo{publisher}{ACM}, \bibinfo{address}{New York, NY, USA},
  \bibinfo{pages}{569--579}.
\newblock
\showISBNx{978-1-4503-5948-1}
\urldef\tempurl%
\url{https://doi.org/10.1145/3242587.3242650}
\showDOI{\tempurl}


\bibitem[\protect\citeauthoryear{Micallef, Adi, and Misra}{Micallef
  et~al\mbox{.}}{2018}]%
        {Login18}
\bibfield{author}{\bibinfo{person}{Nicholas Micallef}, \bibinfo{person}{Erwin
  Adi}, {and} \bibinfo{person}{Gaurav Misra}.} \bibinfo{year}{2018}\natexlab{}.
\newblock \showarticletitle{Investigating Login Features in Smartphone Apps}.
  In \bibinfo{booktitle}{\emph{Proceedings of the 2018 ACM International Joint
  Conference and 2018 International Symposium on Pervasive and Ubiquitous
  Computing and Wearable Computers}} (Singapore, Singapore)
  \emph{(\bibinfo{series}{UbiComp '18})}. \bibinfo{publisher}{Association for
  Computing Machinery}, \bibinfo{address}{New York, NY, USA},
  \bibinfo{pages}{842--851}.
\newblock
\showISBNx{9781450359665}
\urldef\tempurl%
\url{https://doi.org/10.1145/3267305.3274172}
\showDOI{\tempurl}


\bibitem[\protect\citeauthoryear{Mirzaei, Garcia, Bagheri, Sadeghi, and
  Malek}{Mirzaei et~al\mbox{.}}{2016}]%
        {toolTrimdroid2016}
\bibfield{author}{\bibinfo{person}{Nariman Mirzaei}, \bibinfo{person}{Joshua
  Garcia}, \bibinfo{person}{Hamid Bagheri}, \bibinfo{person}{Alireza Sadeghi},
  {and} \bibinfo{person}{Sam Malek}.} \bibinfo{year}{2016}\natexlab{}.
\newblock \showarticletitle{Reducing combinatorics in {GUI} testing of
  {Android} applications}. In \bibinfo{booktitle}{\emph{2016 IEEE/ACM 38th
  International Conference on Software Engineering (ICSE)}}. IEEE,
  \bibinfo{pages}{559--570}.
\newblock


\bibitem[\protect\citeauthoryear{{Moran}, {Bernal-Cárdenas}, {Curcio},
  {Bonett}, and {Poshyvanyk}}{{Moran} et~al\mbox{.}}{2020}]%
        {moran2020-tse}
\bibfield{author}{\bibinfo{person}{K. {Moran}}, \bibinfo{person}{C.
  {Bernal-Cárdenas}}, \bibinfo{person}{M. {Curcio}}, \bibinfo{person}{R.
  {Bonett}}, {and} \bibinfo{person}{D. {Poshyvanyk}}.}
  \bibinfo{year}{2020}\natexlab{}.
\newblock \showarticletitle{Machine Learning-Based Prototyping of Graphical
  User Interfaces for Mobile Apps}.
\newblock \bibinfo{journal}{\emph{IEEE Transactions on Software Engineering}}
  \bibinfo{volume}{46}, \bibinfo{number}{2} (\bibinfo{year}{2020}),
  \bibinfo{pages}{196--221}.
\newblock
\urldef\tempurl%
\url{https://doi.org/10.1109/TSE.2018.2844788}
\showDOI{\tempurl}


\bibitem[\protect\citeauthoryear{Perez and Le}{Perez and Le}{2017}]%
        {perez2017generating}
\bibfield{author}{\bibinfo{person}{Danilo~Dominguez Perez} {and}
  \bibinfo{person}{Wei Le}.} \bibinfo{year}{2017}\natexlab{}.
\newblock \showarticletitle{Generating predicate callback summaries for the
  {Android} framework}. In \bibinfo{booktitle}{\emph{2017 IEEE/ACM 4th
  International Conference on Mobile Software Engineering and Systems
  (MOBILESoft)}}. IEEE, \bibinfo{pages}{68--78}.
\newblock


\bibitem[\protect\citeauthoryear{Pu, Liu, and Wang}{Pu et~al\mbox{.}}{2017}]%
        {webuivision2017}
\bibfield{author}{\bibinfo{person}{Jiachen Pu}, \bibinfo{person}{Jin Liu},
  {and} \bibinfo{person}{Jin Wang}.} \bibinfo{year}{2017}\natexlab{}.
\newblock \showarticletitle{A vision-based approach for deep web form
  extraction}.
\newblock In \bibinfo{booktitle}{\emph{Advanced Multimedia and Ubiquitous
  Engineering}}. \bibinfo{publisher}{Springer}, \bibinfo{pages}{696--702}.
\newblock


\bibitem[\protect\citeauthoryear{Rountev and Yan}{Rountev and Yan}{2014}]%
        {rountev2014static}
\bibfield{author}{\bibinfo{person}{Atanas Rountev} {and}
  \bibinfo{person}{Dacong Yan}.} \bibinfo{year}{2014}\natexlab{}.
\newblock \showarticletitle{Static reference analysis for {GUI} objects in
  Android software}. In \bibinfo{booktitle}{\emph{Proceedings of Annual
  IEEE/ACM International Symposium on Code Generation and Optimization}}.
  \bibinfo{pages}{143--153}.
\newblock


\bibitem[\protect\citeauthoryear{Sahami~Shirazi, Henze, Schmidt, Goldberg,
  Schmidt, and Schmauder}{Sahami~Shirazi et~al\mbox{.}}{2013}]%
        {UIAnalysis13}
\bibfield{author}{\bibinfo{person}{Alireza Sahami~Shirazi},
  \bibinfo{person}{Niels Henze}, \bibinfo{person}{Albrecht Schmidt},
  \bibinfo{person}{Robin Goldberg}, \bibinfo{person}{Benjamin Schmidt}, {and}
  \bibinfo{person}{Hansj\"{o}rg Schmauder}.} \bibinfo{year}{2013}\natexlab{}.
\newblock \showarticletitle{Insights into Layout Patterns of Mobile User
  Interfaces by an Automatic Analysis of {Android} Apps}. In
  \bibinfo{booktitle}{\emph{Proceedings of the 5th ACM SIGCHI Symposium on
  Engineering Interactive Computing Systems}} (London, United Kingdom)
  \emph{(\bibinfo{series}{EICS '13})}. \bibinfo{publisher}{Association for
  Computing Machinery}, \bibinfo{address}{New York, NY, USA},
  \bibinfo{pages}{275--284}.
\newblock
\showISBNx{9781450321389}
\urldef\tempurl%
\url{https://doi.org/10.1145/2494603.2480308}
\showDOI{\tempurl}


\bibitem[\protect\citeauthoryear{Sp{\"a}th, Ali, and Bodden}{Sp{\"a}th
  et~al\mbox{.}}{2019}]%
        {spath2019context}
\bibfield{author}{\bibinfo{person}{Johannes Sp{\"a}th}, \bibinfo{person}{Karim
  Ali}, {and} \bibinfo{person}{Eric Bodden}.} \bibinfo{year}{2019}\natexlab{}.
\newblock \showarticletitle{Context-, flow-, and field-sensitive data-flow
  analysis using synchronized Pushdown systems}.
\newblock \bibinfo{journal}{\emph{Proceedings of the ACM on Programming
  Languages}} \bibinfo{volume}{3}, \bibinfo{number}{POPL}
  (\bibinfo{year}{2019}), \bibinfo{pages}{1--29}.
\newblock


\bibitem[\protect\citeauthoryear{Sp{\"a}th, Nguyen Quang~Do, Ali, and
  Bodden}{Sp{\"a}th et~al\mbox{.}}{2016}]%
        {spath2016boomerang}
\bibfield{author}{\bibinfo{person}{Johannes Sp{\"a}th}, \bibinfo{person}{Lisa
  Nguyen Quang~Do}, \bibinfo{person}{Karim Ali}, {and} \bibinfo{person}{Eric
  Bodden}.} \bibinfo{year}{2016}\natexlab{}.
\newblock \showarticletitle{{Boomerang}: Demand-driven flow-and
  context-sensitive pointer analysis for {Java}}. In
  \bibinfo{booktitle}{\emph{30th European Conference on Object-Oriented
  Programming (ECOOP 2016)}}. Schloss Dagstuhl-Leibniz-Zentrum fuer Informatik.
\newblock


\bibitem[\protect\citeauthoryear{Vall{\'e}e-Rai, Co, Gagnon, Hendren, Lam, and
  Sundaresan}{Vall{\'e}e-Rai et~al\mbox{.}}{1999}]%
        {Soot}
\bibfield{author}{\bibinfo{person}{Raja Vall{\'e}e-Rai}, \bibinfo{person}{Phong
  Co}, \bibinfo{person}{Etienne Gagnon}, \bibinfo{person}{Laurie Hendren},
  \bibinfo{person}{Patrick Lam}, {and} \bibinfo{person}{Vijay Sundaresan}.}
  \bibinfo{year}{1999}\natexlab{}.
\newblock \showarticletitle{Soot -- a {Java} Bytecode Optimization Framework}.
  In \bibinfo{booktitle}{\emph{Proceedings of the 1999 Conference of the Centre
  for Advanced Studies on Collaborative Research}} (Mississauga, Ontario,
  Canada) \emph{(\bibinfo{series}{CASCON '99})}. \bibinfo{publisher}{IBM
  Press}, \bibinfo{pages}{13--}.
\newblock
\urldef\tempurl%
\url{http://dl.acm.org/citation.cfm?id=781995.782008}
\showURL{%
\tempurl}


\bibitem[\protect\citeauthoryear{Wang, Zhang, and Rountev}{Wang
  et~al\mbox{.}}{2016}]%
        {wang2016unsoundness}
\bibfield{author}{\bibinfo{person}{Yan Wang}, \bibinfo{person}{Hailong Zhang},
  {and} \bibinfo{person}{Atanas Rountev}.} \bibinfo{year}{2016}\natexlab{}.
\newblock \showarticletitle{On the unsoundness of static analysis for {Android}
  {GUI}s}. In \bibinfo{booktitle}{\emph{Proceedings of the 5th ACM SIGPLAN
  International Workshop on State Of the Art in Program Analysis}}.
  \bibinfo{pages}{18--23}.
\newblock


\bibitem[\protect\citeauthoryear{Yang, Wu, Zhang, Wang, Swaminathan, Yan, and
  Rountev}{Yang et~al\mbox{.}}{2018}]%
        {gator2018}
\bibfield{author}{\bibinfo{person}{Shengqian Yang}, \bibinfo{person}{Haowei
  Wu}, \bibinfo{person}{Hailong Zhang}, \bibinfo{person}{Yan Wang},
  \bibinfo{person}{Chandrasekar Swaminathan}, \bibinfo{person}{Dacong Yan},
  {and} \bibinfo{person}{Atanas Rountev}.} \bibinfo{year}{2018}\natexlab{}.
\newblock \showarticletitle{Static window transition graphs for {Android}}.
\newblock \bibinfo{journal}{\emph{Automated Software Engineering}}
  \bibinfo{volume}{25}, \bibinfo{number}{4} (\bibinfo{year}{2018}),
  \bibinfo{pages}{833--873}.
\newblock


\bibitem[\protect\citeauthoryear{Yang, Yan, Wu, Wang, and Rountev}{Yang
  et~al\mbox{.}}{2015}]%
        {yang2015static}
\bibfield{author}{\bibinfo{person}{Shengqian Yang}, \bibinfo{person}{Dacong
  Yan}, \bibinfo{person}{Haowei Wu}, \bibinfo{person}{Yan Wang}, {and}
  \bibinfo{person}{Atanas Rountev}.} \bibinfo{year}{2015}\natexlab{}.
\newblock \showarticletitle{Static control-flow analysis of user-driven
  callbacks in Android applications}. In \bibinfo{booktitle}{\emph{2015
  IEEE/ACM 37th IEEE International Conference on Software Engineering}},
  Vol.~\bibinfo{volume}{1}. IEEE, \bibinfo{pages}{89--99}.
\newblock


\bibitem[\protect\citeauthoryear{Zhang, Sui, and Xue}{Zhang
  et~al\mbox{.}}{2018}]%
        {toolCHIME2018}
\bibfield{author}{\bibinfo{person}{Yifei Zhang}, \bibinfo{person}{Yulei Sui},
  {and} \bibinfo{person}{Jingling Xue}.} \bibinfo{year}{2018}\natexlab{}.
\newblock \showarticletitle{Launch-mode-aware context-sensitive activity
  transition analysis}. In \bibinfo{booktitle}{\emph{Proceedings of the 40th
  International Conference on Software Engineering}}.
  \bibinfo{pages}{598--608}.
\newblock


\end{thebibliography}

\end{document}